\documentclass[fleqn,usenatbib]{mnras}
\usepackage{newtxtext,newtxmath}
\usepackage[T1]{fontenc}

\DeclareRobustCommand{\VAN}[3]{#2}
\let\VANthebibliography\thebibliography
\def\thebibliography{\DeclareRobustCommand{\VAN}[3]{##3}\VANthebibliography}

\usepackage{graphicx}	
\usepackage{amsmath}	
\usepackage{comment}

\def\lya{Ly$\alpha$}

\def\lyc{Lyman-continuum}

\def\gammabg{$\Gamma_{\mathrm{bg}}$}
\def\gammaqso{$\Gamma_{\mathrm{qso}}$}
\def\MFP{$\lambda_{\mathrm{MFP}}$}
\def\req{$R_{\mathrm{eq}}$}
\newcommand{\msun}{\mathrm{M}_\odot}
\defcitealias{2021MNRAS.508.1853B}{B21}
\newcommand{\beck}{\citetalias{2021MNRAS.508.1853B}}

\usepackage{xcolor}
\definecolor{notecolor}{rgb}{0.8,0,0}
\definecolor{color2}{rgb}{0.0,0.8,0.0}
\newcommand{\gk}[1]{{\color{notecolor} #1}}

\newcommand{\CII}{[C\,{\small II}]}

\usepackage{tikz}
\definecolor{lime}{HTML}{A6CE39}
\DeclareRobustCommand{\orcidicon}{\hspace{-3mm}
	\begin{tikzpicture}
		\draw[lime, fill=lime] (0,0) 
		circle [radius=0.16] 
		node[white] {\hspace{0.1mm}{\fontfamily{qag}\selectfont \tiny ID}};
		\draw[white, fill=white] (-0.07,0.1) 
		circle [radius=0.01];
	\end{tikzpicture}
	\hspace{-5mm}
}

\foreach \x in {A, ..., Z}{\expandafter\xdef\csname orcid\x\endcsname{\noexpand\href{https://orcid.org/\csname orcidauthor\x\endcsname}
		{\noexpand\orcidicon}}
}

\title[Mean free path]{Robustness of direct measurements of the mean free path of ionizing photons in the epoch of reionization}

\author[Satyavolu et al.]{{Sindhu Satyavolu$^{1} $\thanks{E-mail: sindhu@theory.tifr.res.in}\;\orcidA{}\;\;,
  Girish Kulkarni$^{1}$\;\orcidB{}\;\;,
  Laura C.~Keating$^{2}$\;\orcidC{}\;\;, and Martin G.~Haehnelt$^{3,4}$\;\orcidD{}}\\
  $^{1}$Tata Institute of Fundamental Research, Homi Bhabha Road, Mumbai 400005, India\\
  $^2$Institute for Astronomy, University of Edinburgh, Blackford Hill, Edinburgh, EH9 3HJ, United Kingdom\\
  $^3$Institute of Astronomy, University of Cambridge, Madingley Road, Cambridge CB3 0HA, UK \\
$^4$Kavli Institute of Cosmology, University of Cambridge, Madingley Road, Cambridge CB3 0HA, UK
}

\date{Accepted ---. Received ---; in original form ---}

\pubyear{2022}

\begin{document}
\label{firstpage}
\pagerange{\pageref{firstpage}--\pageref{lastpage}}
\maketitle

\begin{abstract}
Measurements of the mean free path of Lyman-continuum photons in the intergalactic medium during the epoch of reionization can help constrain the nature of the sources as well as sinks of hydrogen-ionizing radiation.  A recent approach to this measurement has been to utilize composite spectra of multiple quasars at $z\sim 6$, and infer the mean free path after correcting the spectra for the presence of quasar proximity zones.  This has revealed not only a steep drop in the mean free path from $z=5$ to $z=6$, but also potentially a mild tension with reionization simulations.  We critically examine such direct measurements of the mean free path for biases due to quasar environment, incomplete reionization, and quasar proximity zones.  Using cosmological radiative transfer simulations of reionization combined with one-dimensional radiative transfer calculations of quasar proximity zones, we find that the bias in the mean free path due to overdensities around quasars is minimal at $z\sim 6$.  Patchiness of reionization at this redshift also does not affect the measurements significantly.  Fitting our model to the data results in a mean free path of $\lambda_{\mathrm{mfp}}=1.49^{+0.47}_{-0.52}$~pMpc at $z=6$, which is consistent with the recent measurements in the literature, indicating robustness with respect to the modelling of quasar proximity zones.  We also compare various ways in which the mean free path has been defined in simulations before the end of reionization.  Overall, our finding is that recent measurements of the mean free path appear to be robust relative to several sources of potential bias.
\end{abstract}


\section{Introduction}

The epoch of reionization marks the era during which the neutral hydrogen in the Universe became largely ionized, most likely due to UV photons from stars in galaxies across the cosmic volume, with potentially an additional minor contribution from quasars.  To understand how reionization happened, we need to know how the photons that escaped from the various sources of radiation interacted with the intergalactic medium (IGM) to ionize the universe. An important characteristic of the propagation of these ionizing photons is determined by their mean free path (MFP), defined as the average distance a photon travels before getting absorbed \citep{1985rpa..book.....R}.  In a homogeneous IGM, the MFP increases with the background UV photoionization rate, as this lowers the IGM opacity.  More generally, the MFP \MFP\ varies with the background photoionization rate \gammabg\ as a power law \citep{2000ApJ...530....1M}, and, furthermore, both of these quantities can be expressed as functions of the cosmological emissivity $\epsilon$ of ionizing radiation.  Knowing the MFP therefore puts constraints on the ionizing sources as well as the sinks of ionizing photons. At higher redshifts, the rate of increase of the MFP with redshift indicates how rapid the progress of reionization is. Measurement of the MFP of hydrogen-ionizing photons is thus important for characterizing reionization. 

During reionization, the sources of radiation carve out regions of ionized hydrogen in the IGM around themselves.  These ionized regions later coalesce, at which point a background of ionizing radiation gets established throughout the Universe \citep{2014ApJ...793...30G, 2015MNRAS.453.3593B}.  This is considered to be the end of reionization and is thought to occur at $z\sim 5.3$ \citep{2019MNRAS.485L..24K, 2022MNRAS.514...55B,2022ApJ...932...76Z}.  At lower redshifts, in the post-reionization Universe, the MFP is dominated by the residual neutral hydrogen systems in the otherwise fully ionized IGM.  These systems retain neutral hydrogen due to self-shielding thanks to their high density.  These appear as high-column-density absorbers of the ionizing radiation background. The corresponding MFP is then set by the average spacing between such absorbers, obtained by counting the number of optically-thick absorbers or Lyman-limit systems (LLS) along the lines of sight in quasar spectra \citep{1994ApJ...427L..13S,2007ApJ...656..666O,2013ApJ...765..137O}. At higher redshifts, before reionization has ended, however, the boundaries of the ionized regions also play an important role in setting the MFP.  At these times, the MFP is the average distance between LLSs or the typical size of ionized regions, depending on which is smaller \citep{2016MNRAS.458..135S, 2017ApJ...851...50M}.

Observationally, measurements of the MFP have been inferred using direct and indirect means.  One way to obtain a direct measurement of the MFP is to use stacked quasar spectra bluewards of rest-frame 912\AA.  The MFP is then computed as the distance at which the effective \lyc\ optical depth becomes unity \citep{2009ApJ...705L.113P,2013ApJ...775...78F,2013ApJ...765..137O,2014MNRAS.445.1745W}.  This has resulted in a measurement of $\lambda_\mathrm{mfp}=10.3\pm 1.6$~pMpc at $z=5.16$, with the MFP increasing as $(1+z)^{-5.4}$ down to $z=4.56$.  At higher redshifts, the MFP can however become comparable or smaller than the size of the proximity zone of the quasars whose spectra are used in the stacks.  This biases the MFP towards lower values \citep{2014MNRAS.445.1745W, 2018MNRAS.473..560D}.  \citet[][B21]{2021MNRAS.508.1853B} reported a measurement of the MFP in which they sought to correct the bias due to the quasar proximity zones by modifying the direct measurement method to account for the excess flux due to the quasar.  The resultant value of the MFP inferred by \citetalias{2021MNRAS.508.1853B} is $0.75^{+0.65}_{-0.45}$~pMpc at $z= 6$.  The MFP thus shows a sharp drop at $z>5$ relative to the $(1+z)^{-5.4}$ behaviour seen at redshifts lower than 5.  Moreover, the value of the MFP at $z=6$ is smaller than the value in several of the latest reionization simulations by a factor of 2 or more \citep{2020MNRAS.491.1736K}.  \citet{2021ApJ...918L..35D} discussed the implications of this tension for the ionizing budget of galaxies to argue that a shorter MFP requires an ionizing emissivity that is up to six times larger than the typically assumed values.  \citet{2021ApJ...917L..37C} found that the short MFP reported by \beck\ is consistent with a late reionization scenario powered by fainter galaxies with a high ionizing photon production efficiency \citep{2022MNRAS.516.3389L,2022MNRAS.512.4909G}.  More recently, \cite{2023ApJ...955..115Z} have measured the MFP for quasars at redshifts between $5.1<z<6$ using the \beck\ method and found the value at redshift 6 to agree with that measured by \beck. They also find the decrease in the MFP to be steeper with increasing redshift, with a sharp drop in MFP by nearly 75$\%$ between $z\sim 5.6$  and 5.9.  Such a sharp decline in the MFP points towards a rapid end to reionization.  Another direct measurement technique is to define the MFP by averaging free paths measured from the distribution of absorbers along individual quasar sightlines \citep{2019A&A...632A..45R,2010ApJ...721.1448S}. This approach suggested a milder evolution of the MFP with redshift, between $3<z<6$, when compared to the direct stacking method. \citet{2021arXiv210812446B} have extended this approach to detect absorption due to LLS in the six lowest-order Lyman-series transitions and put a lower limit based on the average of individual free paths defined this way to be \MFP~$ >0.31$~pMpc at redshift 6, consistent with the measurement of \beck. 

Indirect measurements of the MFP have been obtained by comparing the observed \lya\ opacity of the IGM with that in numerical simulations. The mean and the scatter of the cumulative distribution function (CDF) of the effective \lya\ opacity has been used to simultaneously constrain both the MFP and the photoionization rate for a given emissivity~\citep[e.g., ][and Davies et al. 2023, in preparation]{2023arXiv230402038G,2023MNRAS.521.4056W}.  Overall, the indirect MFP measurements based on the \lya\ opacity \citep{2023arXiv230402038G} are consistent with the direct MFP measured from the quasar stack beyond the Lyman limit, within the error bars \citep{2021MNRAS.508.1853B,2023ApJ...955..115Z}. However, the slope of evolution of the MFP with redshift differs beyond redshift $z\sim 5.5$, with the direct measurements yielding a steeper slope than the indirect measurements. A gradual slope agrees better with the reionization simulations of \citet{2018MNRAS.473..560D}, \citet{2020MNRAS.491.1736K}, and \citet{2021ApJ...917L..37C}, as compared to the steeper slope seen in the models of \citet{2022MNRAS.516.3389L} and \citet{2022MNRAS.512.4909G}.  Possible reasons for the mismatch between the indirect and direct measurements of the MFP could be the assumptions made by the two  methods. \beck\ assume an analytic model to measure the effective optical depth, where the opacity $\kappa$ is proportional to the photoionization rate as $\Gamma^{-\xi}$. In the QSO proximity zone, the total $\Gamma$ is a sum of both \gammabg\ and \gammaqso, the latter predominantly varying with distance from the quasar due to geometric dilution.  \beck\ solve for \gammaqso\ numerically, assigning average parameters for their QSO stack. They thereby keep the background value of the photoionization rate, \gammabg\, fixed while keeping the MFP as a free parameter.  In reality, the photoionization rate and MFP will both depend on the ionizing emissivity and will co-evolve. \beck\ also fix the opacity due to higher-order Lyman series absorption to the values obtained using an optically thin simulation while fitting their model. Roth et al. in prep discuss the effect of this on the inference of \beck.
The slope of the variation of the MFP with redshift is also different between the counting LLS method, indirect inference method and the direct stacking method, the latter showing the steepest evolution. These differences reflect the dependence of the MFP on the nature of absorbing sources, each sensitive to a different measurement technique \citep{2014MNRAS.442.1805I}.

In this work, we critically examine the direct measurement method of \beck\ for possible biases due to (\textit{a}\/) higher cosmological densities around high-redshift quasars, (\textit{b}\/) incomplete reionization at $z>5.3$, and (\textit{c}\/) differences between the structure of quasar proximity zones as computed using the analytical model of \beck\ and that obtained via radiative transfer calculations.  We also investigate the challenge of defining the MFP during the epoch of reionization, when a cosmological UV radiation background is not yet established uniformly.  This paper is structured as follows.  We discuss our data and models in Section~\ref{sec:methods}.   We discuss the definition of the MFP before the end of reionization, and the effect of overdense quasar environments on the MFP in Section~\ref{sec:mfpsims}.  Section~\ref{sec:mfpmock} presents a discussion on the effect of incomplete reionization on the \beck\ measurements by discussing mock measurements of the MFP in our simulations using the \beck\ method.  Finally, we present a new measurement of the MFP using the \beck\ data in Section~\ref{sec:results}, incorporating a radiative transfer model of quasar proximity zones.  We end with a discussion and a summary of our conclusions in Section~\ref{sec:conclusion}.  Our $\Lambda$CDM cosmological model has $\Omega_\mathrm{b}=0.0482$, $\Omega_\mathrm{m}=0.308$, $\Omega_\Lambda=0.692$, $h=0.678$, $n_\mathrm{s}=0.961$, $\sigma_8=0.829$, and $Y_\mathrm{He}=0.24$ \citep{2014A&A...571A..16P}. All distances are in proper (physical) units unless specified otherwise.

\section{Data and Models}\label{sec:methods}

Our strategy in this paper is to use a cosmological radiative transfer (RT) simulation of reionization.  Such a simulation provides us with a realistic model for a partially ionized IGM at $z>5.3$, on top of which we model quasar proximity zones using one-dimensional radiative transfer.  We then use the resultant one-dimensional spectra to examine the MFP.  The cosmological RT simulation used here is described by \citet{2019MNRAS.485L..24K}.  It consists of a cosmological hydrodynamical simulation performed using P-GADGET-3 (which is a modified version of GADGET-2, described by \citealt{2005MNRAS.364.1105S}), that is post-processed with three-dimensional radiative transfer using the ATON code \citep{2008MNRAS.387..295A, 2010ApJ...724..244A}.  The box size of the simulation is 160~cMpc$/h$  with $2048^3$ gas and dark matter particles.  This model is calibrated to reproduce the observed mean \lya\ transmission at $z>5$, and agrees with several high-redshift observations such as the spatial fluctuations in the \lya\ effective opacity \citep{2015MNRAS.447.3402B, 2022MNRAS.514...55B},  the Thomson scattering optical depth to the last scattering surface \citep{2020A&A...641A...6P}, the large-scale radial distribution of galaxies around opaque \lya\ troughs \citep{2020MNRAS.491.1736K,2015MNRAS.447.3402B}, quasar damping wings \citep{2017MNRAS.466.4239G,2019MNRAS.484.5094G,2018ApJ...864..142D,2020ApJ...896...23W}, measurements of the IGM temperature \citep{2020MNRAS.494.5091G}, and the luminosity function and clustering of \lya\ emitters \citep{2018MNRAS.479.2564W,2019MNRAS.485.1350W}. The mid-point of reionization in this simulation is at $z=7$, and reionization ends at $z=5.3$.  At $z=5.95$, the range of halo masses resolved in the simulation is in between $2.32\times10^{8}\;\msun$ and $4.59\times10^{12}\;\msun$ .  We direct the reader to \citet{2019MNRAS.485L..24K} and \citet{2020MNRAS.491.1736K} for further details.

\begin{figure}
  \includegraphics[width=\columnwidth]{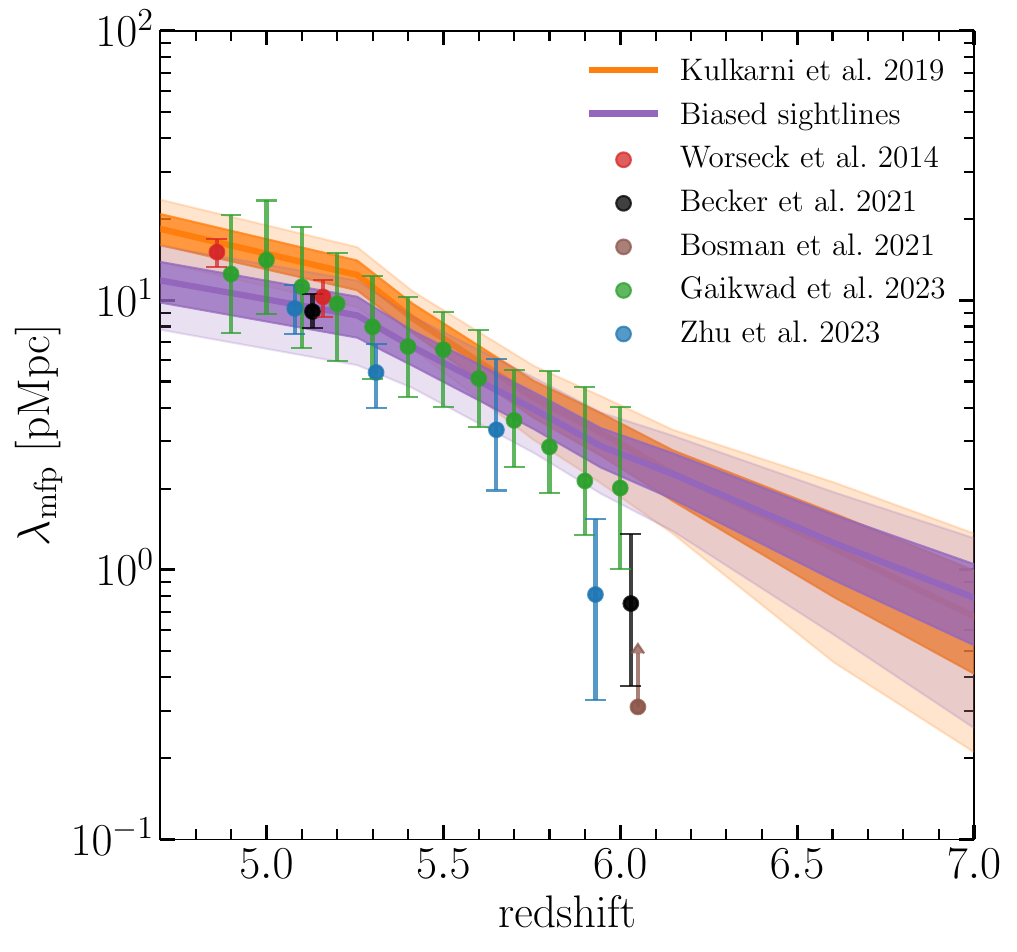}
  \caption{Effect of higher overdensity around high-redshift quasars on the MFP in simulations.  The orange curve and shaded regions show the median value at the one- and two-sigma scatter of the MFP in our reionization model, when the MFP is defined using Equation~(\ref{eqn:mfp1}).  The purple curve and shaded regions show the same but when only sightlines originating in the highest mass halos in the simulation volume are used.  The data points and error bars show various measurements of the MFP. Measurements by \citet{2021MNRAS.508.1853B} and \citet{2021arXiv210812446B} have been displaced in redshift by $\delta z=$ 0.03 and 0.05 respectively, for legibility.}
  \label{fig:mfpsim_v_obs}
\end{figure}

To construct mock spectra including the proximity zone of the QSO, we post-process our simulations with a one-dimensional radiative transfer code that is described by \citet{2023MNRAS.521.3108S}.  All quasars are assigned a magnitude of $M_{1450}=-27.0$, corresponding to the mean magnitude of the quasars in the \beck\ stack.  In our fiducial model, the quasars have a `lightbulb' lightcurve with a lifetime of $10^6$~yr and are placed in halos with masses between $10^{11}\,\msun\lesssim M_{\mathrm{halo}}\lesssim10^{12}\,\msun$.  We assume a broken power-law SED, given by \cite{2015MNRAS.449.4204L}, such that the spectral slopes above and below 912\AA\ are $-0.61$ and $-1.7$ respectively.  We then solve the thermochemistry equations for the hydrogen and helium ionized fractions along a sightline for a given quasar magnitude and lifetime at each redshift. We also simultaneously solve for the gas temperature, which is regulated by photoionization heating, adiabatic cooling, radiative cooling due to recombinations, collisional excitations, Compton cooling due to CMB photons, and Bremsstrahlung.  

For our measurement of the MFP, we use the observed data from \beck.  This is in the form of a composite of quasar spectra with mean redshift $z_\mathrm{qso}=5.97$, constructed with 13 quasar spectra obtained using the Keck/ESI and VLT/XSHOOTER spectrographs. The redshifts used for the quasars in the stack were obtained using the \CII\ line, from the \lya\ halo emission, or from the apparent start of the \lya\ forest absorption.  While selecting quasars in the stack, a minimum cut-off of $S/N\gtrsim$ 20 per 30~km~$\mathrm{s}^{-1}$ was applied for the signal-to-noise ratio near the rest-frame 1285\AA\ in the continuum. The mean brightness of the quasars in the sample is $M_{1450}=-27$, with a range of $-27.8<M_{1450}<-25.7$. The typical resolution for the quasars in the stack is FWHM $\sim45$~km~$\mathrm{s}^{-1}$ for spectra obtained using ESI and FWHM $\sim$25~km~$\mathrm{s}^{-1}$ for spectra obtained using XSHOOTER.
While stacking, the quasar flux is normalised by the median flux between $1270$--$1380$\,\AA, although the normalisation with respect to the intrinsic continuum is kept a free parameter for our analysis, similar to \beck.  In order to study the Lyman-continuum MFP, we focus on the wavelength range of 912--825\AA. 

\section{MFP of hydrogen-ionizing photons before the end of reionization}
\label{sec:mfpsims}

\begin{figure}
  \includegraphics[width=\columnwidth]{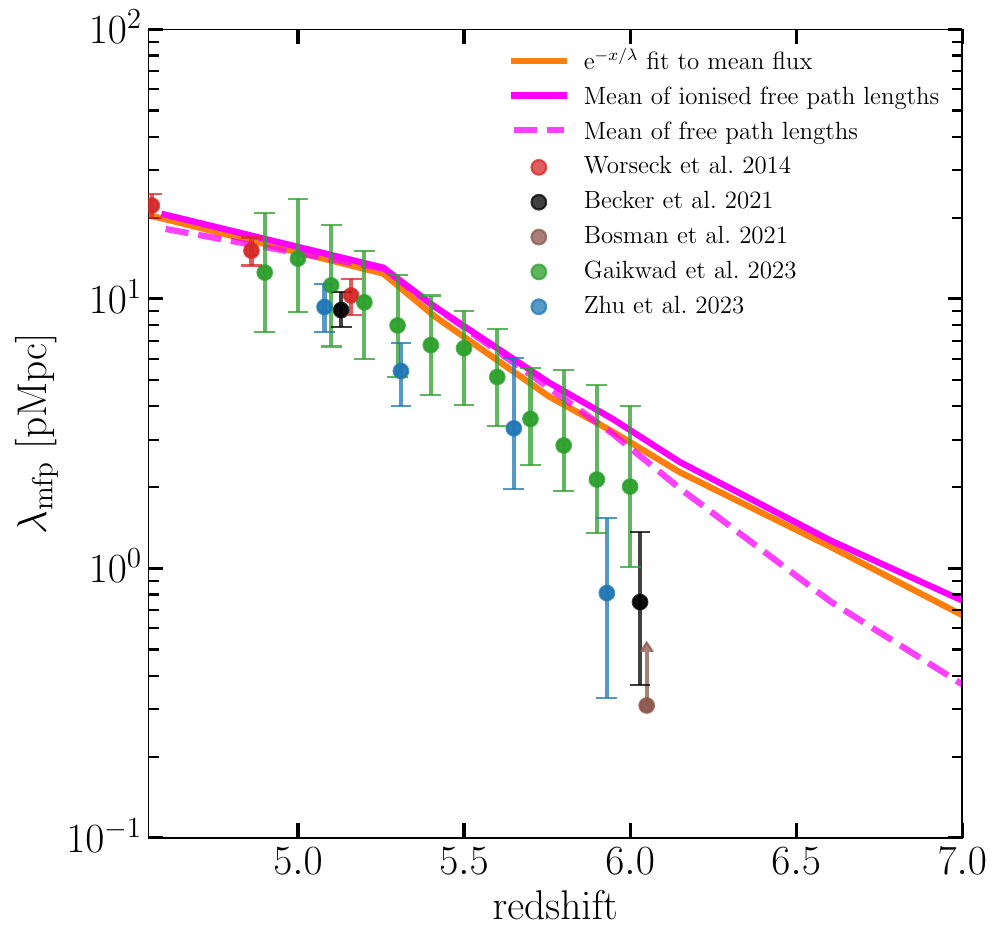}
  \caption{A comparison of three definitions of the MFP in an incompletely reionized IGM.  The solid orange curve shows the MFP obtained from our simulation using Equation~(\ref{eqn:mfp1}). This curve is also shown in Figure~\ref{fig:mfpsim_v_obs}.  The solid pink curve shows the the value of the MFP computed as the mean of the higher-value subpopulation of the bimodal distribution of free-path lengths in the simulation box.  The dashed pink curve shows the MFP when computed as the mean of the full bimodal distribution of the free-path lengths.}
  \label{fig:mfpmethods}
\end{figure}

The post-reionization ($z\lesssim5.5$) MFP measurements are in good agreement with several reionization simulations \citep{2020MNRAS.497..906K,2020ApJ...898..149D,2021ApJ...917L..37C,2022MNRAS.516.3389L}.  At higher redshifts $z\gtrsim 5.5$, when reionization is still believed to be ongoing, both observations and simulations differ in the measured MFP values.  While some simulations agree better with the MFP measured by \beck\ \citep{2022MNRAS.516.3389L,2022MNRAS.512.4909G}, some models \citep[e.g.,][]{2021ApJ...917L..37C}, including ours \citep{2019MNRAS.485L..24K}, agree better with the MFP measured by \citet{2023arXiv230402038G}.  Keeping this in mind, we look at possible biases in the computation of the MFP from the simulations.

We use the standard definition for the MFP \citep{2016MNRAS.463.2583K} where the average \lyc\ transmission across random sightlines in the comoving frame is fit by an exponential with an $e$-folding length scale of \MFP,  
\begin{equation}
  \langle F \rangle = F_{0}\exp\left(-\frac{x}{\lambda_{\mathrm{MFP}}}\right).
  \label{eqn:mfp1}
\end{equation}
Figure~\ref{fig:mfpsim_v_obs} shows the evolution in our model of the MFP defined in this manner.  The orange curve shows the MFP obtained by fitting Equation~(\ref{eqn:mfp1}) to composite spectra of 13 randomly drawn sightlines from our simulation, to match the number of sightlines used in \beck.  In order to obtain the sample variance, we repeat this computation for 10,000 randomly drawn samples. This method for measuring the MFP is analogous to the approach of \citet{2009ApJ...705L.113P}\footnote{We have verified this by fitting our mock quasar stack discussed in Section~\ref{sec:mfpmock} with the model of \citet{2009ApJ...705L.113P}}. The solid curves in Figure~\ref{fig:mfpsim_v_obs} show the median of the resultant distribution of MFP values, with the shaded regions showing the one- and two-sigma scatter.  We see that the disagreement between the MFP measured in this manner from the measured value of \beck\ and \cite{2023ApJ...955..115Z} is more than $2\sigma$ at $z=6$.  Also missing in the simulations is the steep decline in the MFP observed by \beck\ and \cite{2023ApJ...955..115Z} between $z\sim 5.5$ and $z\sim 6$.  We also see that the evolution of the MFP in our model becomes even more gradual once reionization is complete at $z\sim 5.3$.  At these post-reionization redshifts, the optical depth is controlled by the overdense regions responsible for LLSs. 


Direct measurements infer MFP from sightlines towards QSOs, which tend to reside in overdensities \citep[e.g.][]{2023ApJ...951L...4W}. To check for the bias due to large scale structure, we draw random sightlines originating only from halos with the highest masses sampled by our simulation volume, between $10^{12}$ and $10^{13}\mathrm{M}_{\odot}$, otherwise applying Equation~(\ref{eqn:mfp1}) as above.  The resulting median MFP with the one- and two-sigma scatter  is shown in Figure~\ref{fig:mfpsim_v_obs} in purple. The MFP along biased sightlines shows qualitatively similar behavior with redshift when compared to the MFP along random sightlines, but with a slightly shallower slope.  The difference between the two MFPs is insignificant at $z=6$. The random sightlines include sightlines that start from any location, including halos. Any difference seen between the two cases in Figure~\ref{fig:mfpsim_v_obs} is because sightlines originating from massive halos are ionized earlier than other regions in the IGM. Once reionization is complete, the over-density along sightlines originating from halos will also lead to a higher optical depth or lower MFP when compared to the measurement of random sightlines.  We see in Figure~\ref{fig:mfpsim_v_obs} that this biased MFP is in better agreement with the measurement by \cite{2023ApJ...955..115Z} and \beck\ than that by \citet{2014MNRAS.445.1745W} at $z\lesssim 5.5$. We discuss this further in Section~\ref{sec:mfpmock} below. 

Another potential source of bias in the MFP as formally defined by Equation~(\ref{eqn:mfp1}) is that this exponential attenuation assumes a spatially constant opacity.  In reality, reionization is not yet complete at $z=6$.  Consequently, the \lyc\ opacity of the IGM has large spatial variations.  An alternative definition of the MFP that addresses this complexity is one that obtains a distribution of the free-path lengths of photons in a simulation box and computes its mean~\citep{2018MNRAS.478.5123R}.  We implement this in our simulation by following \citet{2018MNRAS.478.5123R} and measuring the free path as the distance at which the \lyc\ optical depth along the sightline becomes unity.  The distribution of such free-path lengths is bimodal \gk{(see Appendix~\ref{sec:appendixa})}.  One part of the distribution describes the mean free path in neutral regions that are on average not overdense.  The sightlines with larger free paths on the other hand also include those which encounter neutral islands along the path.  The fraction of sightlines with small free paths becomes smaller than those with larger free paths at lower redshifts, until the free path distribution becomes unimodal Gaussian.  We measure the MFP as the average of the dominant part of the bimodel distribution, which represents the ionized IGM.  The MFP defined in this manner is shown by the solid pink curve in Figure~\ref{fig:mfpmethods}.  We see that this results in a value of the MFP that is very close to the value obtained by using Equation~(\ref{eqn:mfp1}), shown by the orange curve in Figure~\ref{fig:mfpmethods}.

In their CoDa III simulation, \cite{2022MNRAS.516.3389L} compute the MFP using a similar free-path length method. The free-path lengths are defined in this case by doing an exponential fit to the flux along individual sightlines, which they find to be similar to the free paths measured following \cite{2018MNRAS.478.5123R}. 
The dashed pink curve in Figure~\ref{fig:mfpmethods} shows the result from our simulation when we average over all free paths, similar to one of the approaches used in \cite{2022MNRAS.516.3389L} to measure the MFP.  As expected from the shape of the distribution of the free-path lengths, the MFP is now biased towards lower values.  We find that this bias is small at $z\lesssim 6$.  This could be a reflection of the inadequate spatial resolution of our simulation, due to which a large number of free paths become smaller than our cell size.

\begin{figure}
  \includegraphics[width=\columnwidth]{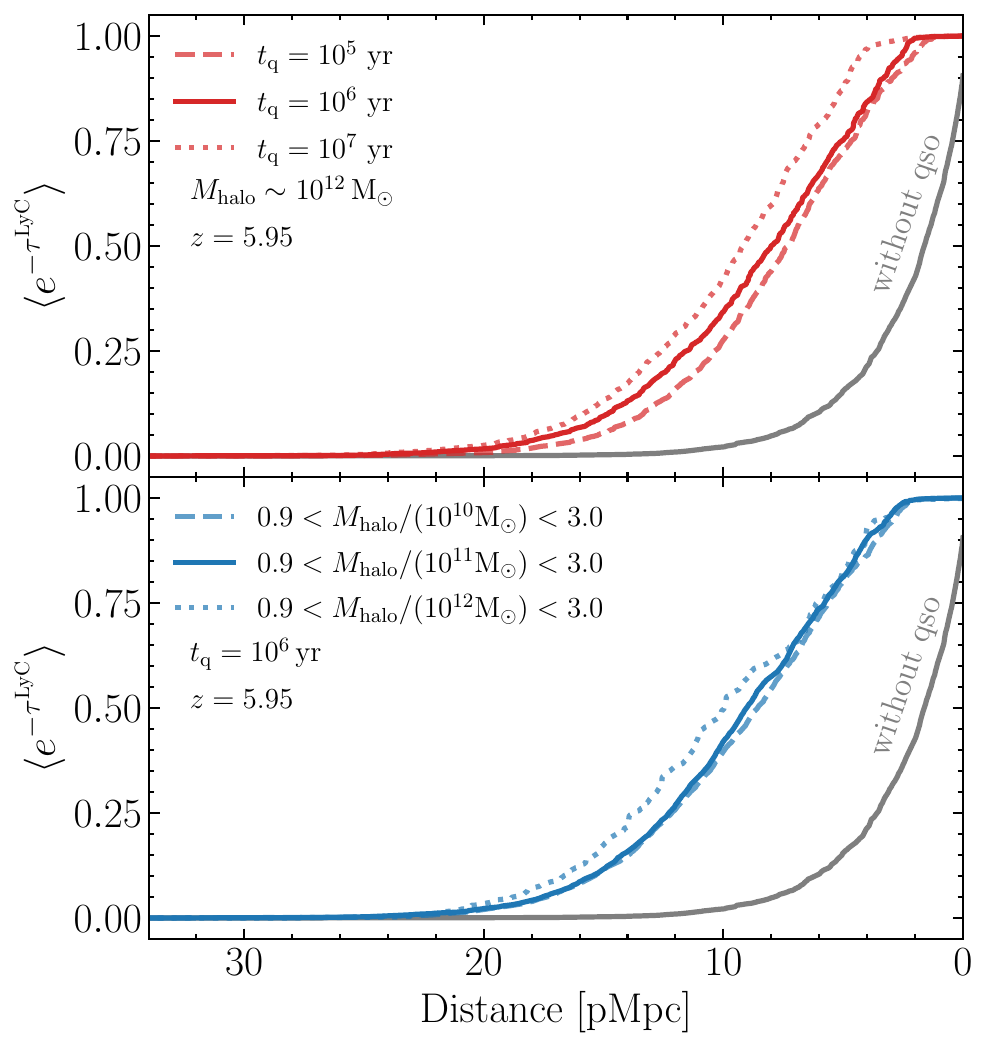}
  \caption{Effect of quasar lifetime (top panel) and host halo mass (bottom panel) on the \lyc\ composite spectrum of 1000 sightlines in our simulation.  In both cases, the grey curve shows the transmission without the effect of the quasar proximity zone. The volume averaged neutral hydrogen fraction in our simulation at $z=5.95$ is $\langle x_{\rm{HI}}\rangle=0.13.$}
  \label{fig:stacktqhm}
\end{figure}

\section{Effect of incomplete reionization on the B21 measurement}
\label{sec:mfpmock}
We have seen above that neither the large-scale-structure bias nor the variations in the definition of the MFP cause a significant change in the value of the MFP in our simulation at $z=6$.  We now investigate the effect of the residual neutral hydrogen `islands' in the IGM at this redshift on the \beck\ measurement of the MFP.  Given the good agreement of our simulations with the measurements of the MFP by \citet{2014MNRAS.445.1745W}, \citet{2023arXiv230402038G} and \citet{2021arXiv210812446B}, it is important to consider if the absence of patchy reionization in the models of \beck\ and \cite{2023arXiv230804614Z} could potentially bias their measurements.  In order to do this, we construct mock data from our simulation and apply the \beck\ method to it.  We then compare the resultant measurement of the MFP with the `true' value of the MFP in the simulation given by Equation~\ref{eqn:mfp1}.  A similar test was also performed by \beck\ themselves.  However, the mock spectra used in their test were created from numerical simulations where reionization is assumed as instantaneous from a uniform ionizing background and post-processed to include fluctuations following the approach of \cite{2016MNRAS.460.1328D}.

\begin{figure}
  \includegraphics[width=\columnwidth]{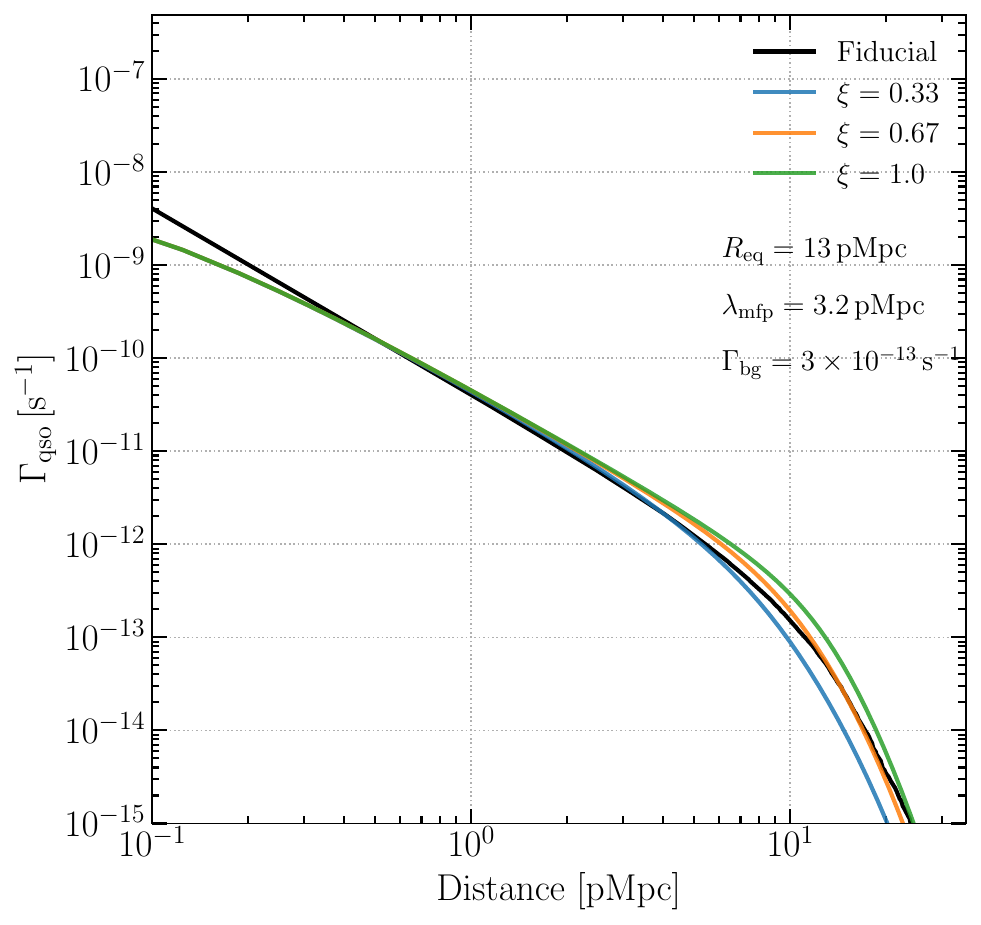}
  \caption{HI photoionization rate due to QSO: The black curve shows the average $\Gamma_{\mathrm{qso}}$ for QSOs with lifetimes ranging between $10^5<t_{\mathrm{q}}<10^7$ yr with magnitude $M_{1450}=-27$ at redshift $z=5.95$, computed using our 1D radiative transfer simulation. The green, orange and blue curves show the average $\Gamma_{\mathrm{qso}}$ computed analytically for the same QSO magnitude at redshift 6, for $\xi=0.33,0.67$, and 1.0 respectively. The value of \req\ is computed using $\Gamma_{\mathrm{bg}}=3\times10^{-13}\mathrm{s}^{-1}$.}
  \label{fig:gammaqso}
\end{figure}

\subsection{The \beck\ method}
We begin by briefly reviewing the \beck\ method for measuring the MFP. The effective \lyc\ optical depth for a photon that is emitted at $z_{\mathrm{qso}}$ and redshifts to the Lyman limit at $z_{912}$ is given by 
\begin{equation}
	\tau^{\mathrm{LL}}_{\mathrm{eff}} (z_{912})=  \frac{c}{H_{0}\Omega_{\mathrm{m}}^{1/2}}(1+z)^{2.75}\int_{z_{912}}^{z_{\mathrm{qso}}} \kappa_{912}(z') (1+z')^{-5.25}\mathrm{d}z',
\end{equation}
where $\kappa_{912}$ is the opacity to 912\,\AA\ photons.  The dependence on redshift is as follows: the exponent $2.75$ is a result of the dependence of cross-section on frequency as $\sigma^{\rm{LL}}_{\nu} \propto \nu^{-2.75}$, while the exponent $-5.25$ comes from the conversion of comoving distance to redshift as $r\propto aH\propto (1+z)^{-1}\times(1+z)^{-3/2}$ in the matter-dominated era. The opacity $\kappa_{912}$ is assumed to scale with the photoionization rate as a power law, so that at a distance $r$ from the quasar,
\begin{equation}
	\kappa_{912} (r) = \kappa_{912}^{\mathrm{bg}}\left(\frac{\Gamma_{\mathrm{tot}}(r)}{\Gamma_{\mathrm{bg}}}\right)^{-\xi},
\end{equation}
where $\kappa_{912}^{\mathrm{bg}}$ is the opacity to the ionized background and $\Gamma_{\mathrm{tot}}(r)=\Gamma_{\mathrm{qso}}(r)+\Gamma_{\mathrm{bg}}(r)$ is the sum of the photoionization rate due to the QSO and the local background photoionization rate. The opacity and photoionization rate are related to each other through their mutual dependence on the shape and number density of neutral gas absorbers. This information is parameterised using the power-law index $\xi$, which has been studied using analytic arguments as well as numerical simulations \citep{2016MNRAS.455.1385M,2011ApJ...743...82M}. We discuss this parameter in detail in Section~\ref{sec:mfpmock}.
Equation~\ref{eq:gammaqso} is applied to a stack of QSO spectra, so the terms in the equation are all averaged quantities, and the average of the local background photoionization is assumed to be uniform and equal to $\Gamma_{\mathrm{bg}}$. 
The average photoionization rate due to all quasars at a location $r$ is computed by iteratively solving for
\begin{equation}
	\Gamma_{\mathrm{qso}} (r+\delta r)= \Gamma_{\mathrm{qso}}(r)\left(\frac{r+\delta r}{r}\right)^{-2}e^{-\kappa_{912}\delta r},
	\label{eq:gammaqso}
\end{equation}
with the initial condition
\begin{equation}
	\Gamma_{\mathrm{qso}}(\delta r) = \Gamma_{\mathrm{bg}}\left(\frac{\delta r}{R_{\mathrm{eq}}}\right)^{-2}.
\end{equation}
Here, \req\ is the distance at which the photoionization rate due to the quasar's radiation is equal to the background photoionization rate in the absence of any absorption. \req\ depends on the QSO magnitude and spectrum in addition to the background photoionization rate.  The free parameters in this model are therefore, $\kappa_{912}^{\mathrm{bg}}$, \req, $\xi$ and $\Gamma_{\mathrm{bg}}$.  The mean flux of the stack is fit by the above model of opacity and the MFP \MFP\ is inferred as the distance from the quasar at which $\tau^{\mathrm{LL}}_{\mathrm{eff}} (z_{912})$ becomes equal to 1.

\begin{figure}
  \includegraphics[width=\columnwidth]{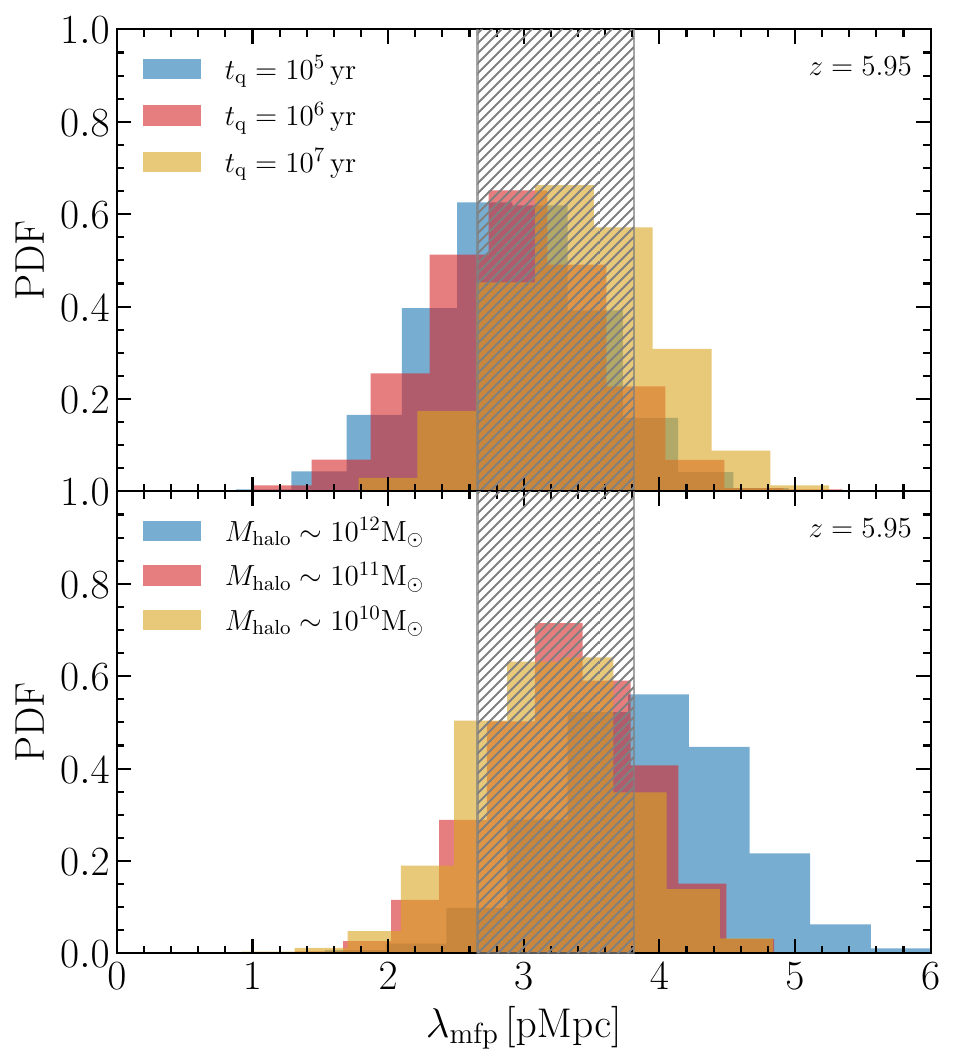}
  \caption{Distribution of the MFP obtained using 10000 draws of 13-sightline stacks from our simulation for different quasar ages (top panel), and different quasar host halo mass (bottom panel).  In each case, the MFP is determined using the \beck\ method, where \req\ and $\Gamma_{\mathrm{bg}}$ are chosen from our simulation and $\xi=0.67$.  As a comparison, the hatched region shows the 68\% scatter in the MFP in the simulation, obtained using Equation~(\ref{eqn:mfp1}).}
  \label{fig:mfppdf_tqhm}
\end{figure}

\subsection{Mock data}

To construct stacks of mock spectra at redshift 6, we consider QSOs with magnitude of $M_{1450}=-27$ with a broken power law spectral profile for the specific luminosity as in \citet{2015MNRAS.449.4204L},
\begin{equation}
	L_\nu\propto\begin{cases}
		\nu^{-0.61} & \text{if}~\lambda\ge 912~\text{\AA},\\
		\nu^{-1.70} & \text{if}~\lambda<912~\text{\AA}.                
	\end{cases}
\end{equation}
Although it is not a deciding factor, we choose the magnitude of our mock stack to be of the value same as the mean magnitude of quasars in the B21 stack at redshift 6. The power law index values are slightly different from those assumed in \beck\ ($-0.5$ and $-1.5$ for wavelengths greater than and less than 912~\AA\ respectively). 

After placing QSOs in halos, we post-process the sightlines with our 1D radiative transfer code as described in Section~\ref{sec:methods} to obtain the distribution of the neutral hydrogen density.  Using this, we compute the ionizing optical depth along a sightline as, 
\begin{equation}
 \tau^{\mathrm{LL}}(r) = \sum_{r} n_{\mathrm{HI}}\sigma_{\mathrm{HI}}^{\mathrm{LL}}\mathrm{d}r,
\end{equation}
where $\sigma_{\mathrm{HI}}^\mathrm{LL}=6.3\times10^{-22}\mathrm{m}^{2}$ is the hydrogen ionization cross-section.
 Since Equation~\ref{eq:gammaqso} does not explicitly assume the QSO lifetime or the host halo mass, we randomize over them in our stack, with halo masses between $10^{10}\msun\lesssim M_{\mathrm{halo}}\lesssim 10^{12}\msun$ and lifetimes between $10^5<t_{\mathrm{q}}<10^7$~Myr. Each of our composite stacks comprises 13 sightlines.
 
Figure~\ref{fig:stacktqhm} shows composite stacks obtained in this manner with and without quasars for the above range of quasar lifetimes and host halo masses.  The top panel shows the variation of the mean continuum flux along quasar sightlines originating from $\sim10^{12}\msun$ halos and having lifetimes between $10^5<t_{\mathrm{q}}<10^7$~Myr. The mean flux computed along sightlines drawn at random directions is shown in grey. Once the quasar is on, the opacity near the quasar decreases as the quasar emits ionizing photons that reduce the neutral hydrogen density. The mean flux thus increases with increase in the quasar lifetime. While the decrease in flux is gradual when the quasar turns on in a uniformly ionized medium, we notice that the flux fall off is steeper in our patchy model. The bottom panel shows the variation of the mean continuum flux when the quasar lifetime is fixed to be $t_{\mathrm{q}}=10^6$~yr and the quasar host halo masses are varied to be in three ranges between $10^{10}\msun\lesssim M_{\mathrm{halo}}\lesssim 10^{12}\msun$. In our simulation, more massive halos are ionized earlier and therefore the mean flux is slightly higher along these sightlines. Figure~\ref{fig:stacktqhm} shows that the mean flux changes almost doubles in the presence of the quasar, while changing the lifetime or host halo masses of the quasars by an order of magnitude changes the mean flux by less than around 10\%. For the purpose of recovery of the MFP using the \beck\ method, we do not include the absorption due to higher order Lyman series transitions in the mock spectra. We correspondingly fit the mock spectra without including the contribution from the Lyman series opacity, as shown in Equation~\ref{eq:fluxfit}. We also do not add a zero-point correction to our mock stack and hence do not include this parameter while fitting using Equation~\ref{eq:fluxfit}.

\subsection{Choice of $R_{\mathrm{eq}}$, $\Gamma_{\mathrm{bg}}$ and $\xi$}

Following \beck\, we fit the mock spectra by an exponential
\begin{equation}
	\langle F\rangle = F_{0} \exp\left({-\tau^{\mathrm{LL}}_{\mathrm{eff}} } \right),
	\label{eq:fluxfit}
\end{equation} 
 with $\kappa_{912}^{\mathrm{bg}}$ as free parameter.  The other parameters are \req, $\Gamma_{\mathrm{bg}}$ and $\xi$, which we discuss below. We chose to keep these parameters fixed, as \beck\ did for their nominal measurement. We have inspected the results while keeping $\xi$ as a free parameter, and found them to be consistent with what was reported in \citet{2023ApJ...955..115Z}. We leave self-consistent parametrization of \gammabg\ with $\kappa_{912}^{\mathrm{bg}}$ for future work.
 
The distance \req\ is given by \citet{2011MNRAS.412.2543C}
 \begin{equation}
 	R_{\mathrm{eq}}  = \left[\frac{L_{912}\sigma_{0}}{8\pi h^2\Gamma_{\mathrm{bg}}(\alpha_{\nu}^{\mathrm{ion}}+2.75)}\right]^{1/2}.
 	\label{eq:req}
 \end{equation}
 The analytic expression for \req\ is computed under the approximation that the quasar photoionization rate falls as $1/r^2$, where $r$ is the distance from the source. This equation can be used to further compute $\Gamma_{\mathrm{qso}}$ in the presence of absorption by iterating over Equation~\ref{eq:gammaqso}. For the mock stacks, we chose a value of \req\ = 13~pMpc, for $L_{912}$ corresponding to magnitude of $M_{1450}=-27.0$ and \gammabg\ of $3\times10^{-13}\,\mathrm{s}^{-1}$ as discussed below.

 \begin{figure}
   \includegraphics[width=\columnwidth]{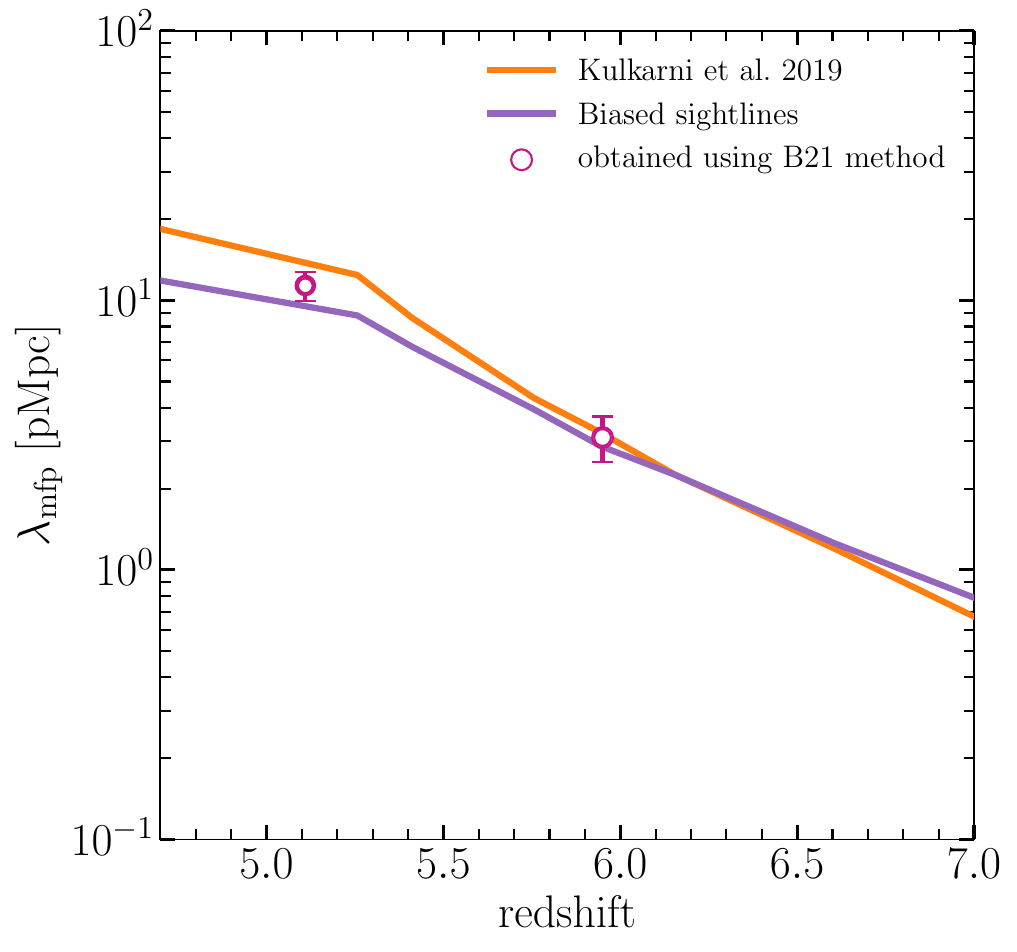}
   \caption{Open circles show the MFP inferred using the \beck\ method from a mock composite spectrum drawn from our simulation.  The curves shows the MFP in the simulation, obtained using Equation~(\ref{eqn:mfp1}), using sightlines starting from randomly chosen halos (orange curve), and sightlines starting from only the most massive halos (purple curve).   When we use the \beck\ method for measuring the MFP, we use $\xi=0.67$ and use values of \req\ and $\Gamma_{\mathrm{bg}}$ derived from the ionized regions in the simulation.}
  \label{fig:mfpsim_b21}
\end{figure}

\beck\ chose a value of $3\times10^{-13}\,\mathrm{s}^{-1}$ for $\Gamma_{\mathrm{bg}}$ in their analysis. 
The background photoionization rate in our simulations is inhomogeneous. In ionized regions, we find the average $\Gamma_{\mathrm{bg}}$ to be around  $3\times10^{-13}\,\mathrm{s}^{-1}$ at redshift $z\sim6$. However,  the average value of $\Gamma_{\mathrm{bg}}$ is $10^{-13}\,\mathrm{s}^{-1}$, almost 3 times smaller than the optically thin value assumed by \beck. Given that at redshift $z\sim6$, the volume averaged ionized hydrogen fraction is close to $80\%$, it would suffice to use this value of background photoionization rate to compute $\Gamma_{\mathrm{qso}}$. We note that \cite{2023arXiv230804614Z} use a different $\Gamma_{\mathrm{bg}}$ of $1.5\times10^{-13}\mathrm{s}^{-1}$  at $z\sim 6$ from \cite{2023arXiv230402038G}.  They find that at $z\sim5.93$, assuming a value of $3\times10^{-13}\,\mathrm{s}^{-1}$ for $\Gamma_{\mathrm{bg}}$ increases their mean free path from the measured value of 0.81~pMpc to $\sim$1~pMpc.
The parameter $\xi$ encodes the nature of the density of the absorbers that set the local MFP \citep{2005MNRAS.363.1031F}. For an isothermal absorber, the theoretical prediction is $\xi=2/3$, while the scaling inferred from simulations is dependent on the self-shielding systems and can range between $0.33$ and $1.0$ \citep{2011ApJ...743...82M,2020ApJ...898..149D}. For $\xi$, we use the nominal parameter from \beck\, where it is assumed to be 0.67 based on the arguments presented in~\cite{2000ApJ...530....1M}. \texttt{\texttt{}}In Figure~\ref{fig:gammaqso}, we compute the photoionization rate using Equation~\ref{eq:gammaqso} for a quasar at redshift 6 and magnitude $M_{1450}=-27$. \req\ was computed using Equation~\ref{eq:req}, assuming $\Gamma_{\mathrm{bg}}$ of $\sim 3\times10^{-13}\,\mathrm{s}^{-1}$. The resulting $\Gamma_{\mathrm{qso}}$ is compared to the $\Gamma_{\mathrm{qso}}$ from our simulations averaged over 1000 lines of sight to the QSO. The analytic $\Gamma_{\mathrm{qso}}$ is in agreement with our simulated value within 25\%, for a $\xi$ value of 0.67. The analytic method however, under-estimates $\Gamma_{\mathrm{qso}}$ by almost two orders of magnitude close to the quasar for all values of $\xi$. Keeping $\xi$ as a free parameter can change the MFP measurement by up to a factor of 2, as has been discussed in  \citet{2023arXiv230804614Z}. Roth et al. in prep discuss that  keeping $\xi$ as a free parameter yields a better fit to their mock spectra in the presence of neutral islands. They however find that such a fit might result in a MFP that is biased high with respect to the true value in their simulation.

\subsection{Recovery of the `true' MFP using the \beck\ method}

We use the \beck\ method to measure the MFP in our simulation using the mock data generated using the methods discussed in Section~\ref{sec:mfpmock}. We find that the resulting MFP agrees with the MFP of our simulation measured in \citet{2019MNRAS.485L..24K}. This shows that the MFP measured using the \beck\ method is not biased due to the residual neutral hydrogen islands, quasar lifetimes or due to the overdensities around quasars.

In Figure~\ref{fig:mfppdf_tqhm}, we show the distribution of MFP measured in our simulations along 10000 stacks of 13 QSOs each. In the fiducial model, the MFP is measured along random sightlines without a QSO by simply fitting an exponential to the average flux as discussed in Section~\ref{sec:mfpsims}. To compute MFP from mock stacks including a QSO, we use the \beck\ method with parameters discussed in the previous subsection. We consider mock stacks with different QSO properties, varying QSO lifetimes and varying host halo masses. The resulting distribution of MFP for 10000 stacks each of 13 QSOs for different QSO lifetimes and host halo masses in shown in the top and bottom panels of Figure~\ref{fig:mfppdf_tqhm}, respectively.  We find that irrespective of the quasar lifetime, we are able to recover the MFP from the mock stacks using the \beck\ method reasonably well, up to within $\sim$ 10\% of our fiducial value. Similarly, the MFP computed from mock stacks with lower mass halos is in good agreement with the fiducial MFP. If we use a quasar stack constructed along sightlines originating in the heaviest mass halos with $10^{11}\,\msun\lesssim M_{\mathrm{halo}}\lesssim 10^{12}\,\msun$, the measured MFP distribution is offset from the fiducial value by $\sim 25\%$.

\begin{figure}
  \includegraphics[width=\columnwidth]{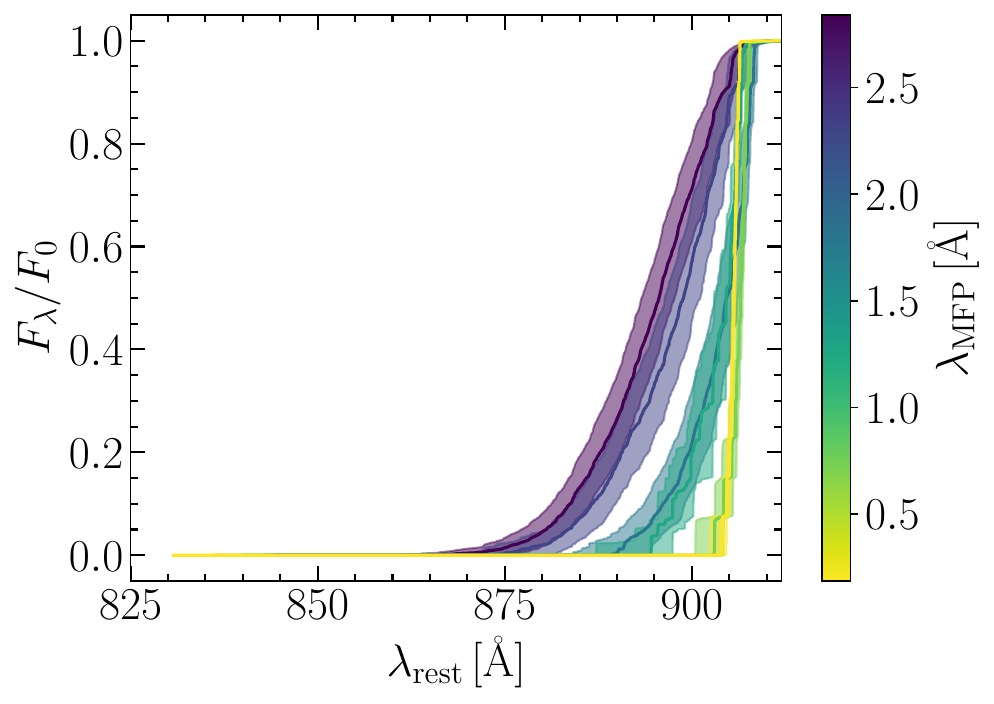}
  \caption{Stacked quasar spectra bluewards of rest-frame 912\AA\ in our model for a range of values of the MFP at $z=6$.  In each case, the solid curve shows the composite of 13 randomly chosen sightlines from the simulation box.  The shaded region shows the 68\% scatter across 10,000 such samples.}
  \label{fig:mockstack_lyc_mfp}
\end{figure}

We show the MFP measured from the 10,000 mock stacks for nominal values of quasar lifetime ($10^6\,\mathrm{yr}$) and host halo mass ($\sim 10^{12}\,\msun$) at redshifts $z=5.1$ and 6 in Figure~\ref{fig:mfpsim_b21}. Also shown are the MFPs along random sightlines and biased sightlines originating in halos but without a QSO, as discussed in Section~\ref{sec:mfpsims}. The MFP recovered from our mock stacks using the \beck\ method matches better with the MFP measured along biased sightlines, which is particularly evident at post-reionization redshifts. This would suggest that the analytic computation of $\Gamma_{\mathrm{qso}}$ is robust in accounting for the QSO ionization flux, but is biased to measure the MFP along sightlines originating in overdensities. Note however that at higher redshifts, this bias becomes negligible. Another concern with the analytic model is the uncertainties introduced by the assumptions about the free parameters in the model, \req, $\xi$ and $\Gamma_{\mathrm{bg}}$. \cite{2023arXiv230804614Z} have shown that increasing or decreasing the value of $\xi$ can half or double the MFP estimates relative to the estimates with nominal values. This raises a need for a better constrained value of the MFP independent of the choice of $\xi$. For this purpose, we use simulated models as discussed in the following section.

\section{A direct measurement of the MFP using radiative transfer models}\label{sec:results}

We now proceed to analyse the composite spectrum of \beck\ using our radiative transfer models of patchy reionization and quasar proximity zones.   Figure~\ref{fig:mockstack_lyc_mfp} shows the composite spectra from 13 randomly drawn sightlines along a QSO with magnitude $M_{1450}=-27.0$ and lifetime of 1~Myr in our simulation for a range of values of the mean free path.  As our reionization simulation is calibrated to the observed \lya\ mean transmission, there is no freedom to change the mean free path at a given redshift.  Instead, to construct Figure~\ref{fig:mockstack_lyc_mfp}, we use the values of the ionized hydrogen fraction, the gas density, and temperature from different redshifts, in each case scaling the densities by $(1+z)^3$ to $z=6$.

\begin{figure}
  \includegraphics[width=\columnwidth]{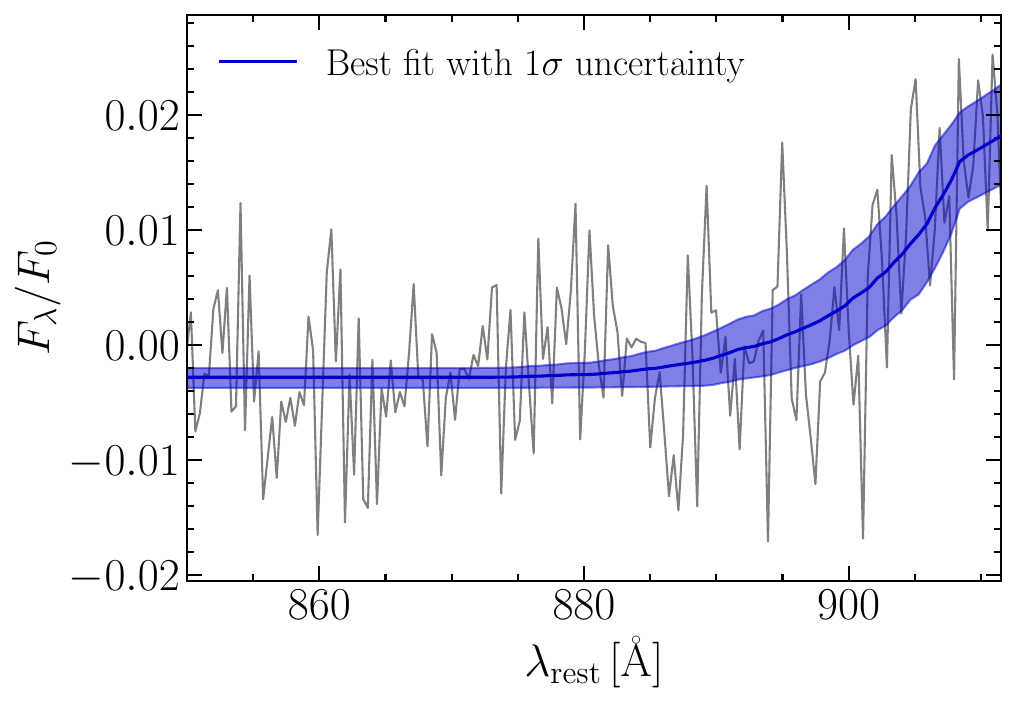}  
  \caption{Our best-fit model for the \beck\ composite spectrum, with the 1$\sigma$ uncertainty indicated by the shaded region.}
  \label{fig:bestfit}
\end{figure}

In order to fit the model stacks in Figure~\ref{fig:mockstack_lyc_mfp} to the data, two modifications need to be made. First, we need to account for the absorption due to higher-order Lyman-series transitions that will cause absorption bluewards of the rest-frame quasar Lyman-limit. Second, to match the observed flux, we need to consider the absorption due to the Lyman-series and continuum photons of the intrinsic QSO spectrum. Following \beck\, we assume the intrinsic quasar spectrum is a power law with~$f_{\mathrm{\lambda}}^{\mathrm{SED}}=f_{912}(\lambda/912\,\textup{\AA})^{-\alpha_{\mathrm{ion}}}$, and keep $f_{912}$ as a free parameter.

We include the foreground absorption due to higher-order Lyman-series transitions as follows.  At a given $\lambda_{\mathrm{obs}}$, the absorption due to transition of the jth Lyman series line will happen due to an absorber at a redshift~$z_{j}=\lambda_{\mathrm{obs}}/\lambda_{j} - 1$. The effective Lyman series optical depth due to all transitions is then, 
\begin{equation}
  \tau_{\mathrm{eff}}^{\mathrm{Lyman}}(\lambda_{\mathrm{obs}})  = \sum_{j} \tau_{\mathrm{eff}}^{j}(z_{j})
\end{equation}
 
 To obtain the optical depth due to the jth transition, we rescale the corresponding \lya\ optical depths at $z_{j}$ by the  product of their oscillator strength and rest-frame wavelength such that~$\tau_{j}=\tau_{\alpha}f_{j}\lambda_{j}/f_{\alpha}\lambda_{\alpha}$.  This scaling is valid in the optically thin regime corresponding to most of the redshifts of our interest, while in the damping wings the optical depths scale roughly as square of oscillator strengths \citep{2015ApJ...799..179M}. 
We use the post-processed sightlines to compute the \lya\ optical depths in the presence of a quasar. The optical depths outside the proximity zone were assumed to be from the 40-2048 Sherwood simulation \citep{2017MNRAS.464..897B}, rescaled to the match the observed mean flux at low $z$ as discussed in \beck. 
From the mean  flux at $z_{j}$, computed as negative exponential of  $\tau_{j}$,  the corresponding effective optical depth  is obtained  as $\tau_{\mathrm{eff}}^{j}=-\mathrm{log}\langle F \rangle$.
Including the absorption from Lyman-limit and 38 higher order Lyman-series terms, the observed flux is
\begin{equation}
  F_\lambda = f_{912}\left(\frac{\lambda}{912\,\textup{\AA} }\right)^{-\alpha_{\mathrm{ion}}} \exp(-\tau_{\mathrm{eff}}^{\mathrm{Lyman}})\exp(-\tau_{\mathrm{eff}}^{\mathrm{LyC}}) + f_0
\end{equation}
where $f_0$ is the zero-point correction factor, that accounts for the uncertainty in the estimated zero-point of the data in \beck. We fit our models to the data by sampling the posterior distribution in a Bayesian manner. We use the likelihood
\begin{equation}
   \mathcal{L}(\bf{d}|\theta)=\frac{1}{\sqrt{(2\pi)^n\,\mathrm{det}\,\bf{C}}}\exp\left[-\frac{1}{2}\left(\bf{m-d}\right)^\mathrm{T} \bf{C^{-1}} \left(\rm{\bf{m-d}}\right)\right],
   \label{eq:likelihood}
\end{equation}
where the column matrices $\bf{m}$ and $\bf{d}$ denote the model and data vectors, respectively, both with $n$ elements corresponding to the number of wavelength bins used. The covariance matrix $\bf{C}$ is obtained as follows.  The data covariance matrix $\bf{C_{\rm{data}}}$ over the wavelength range of our interest (912--850~\AA) is computed from the data by \beck, using bootstrap realisations of the mean flux. Due to the limited data set of 13 quasars, this matrix is noisy and may underestimate the sample variance. We therefore compute a separate bootstrapped model covariance matrix for each of the model parameters using 30,000 simulated stacks and obtain a model correlation matrix $r$. We then regularise the data covariance matrix $\bf{C_{\rm{data}}}$ using the model correlation matrix $r$ to obtain the covariance matrix $\bf{C}$ used in Equation~(\ref{eq:likelihood}) at each point in the parameter space as
\begin{equation}
    \textbf{C}^{ij} = \frac{r^{ij}_{\rm{sim}}}{\textbf{C}_{\rm{data}}^{ii}\textbf{C}_{\rm{data}}^{jj}},
\end{equation}
where
\begin{equation}
  r^{ij}_{\rm{sim}} = \frac{\textbf{C}^{ij}_{\rm{sim}}}{\textbf{C}_{\rm{sim}}^{ii}\textbf{C}_{\rm{sim}}^{jj}}.
\end{equation}
See applications by \citet{2006ApJ...638...27L,2013PhRvD..88d3502V,2017PhRvD..96b3522I} for a discussion of this approach. 

\begin{figure}
  \includegraphics[width=\columnwidth]{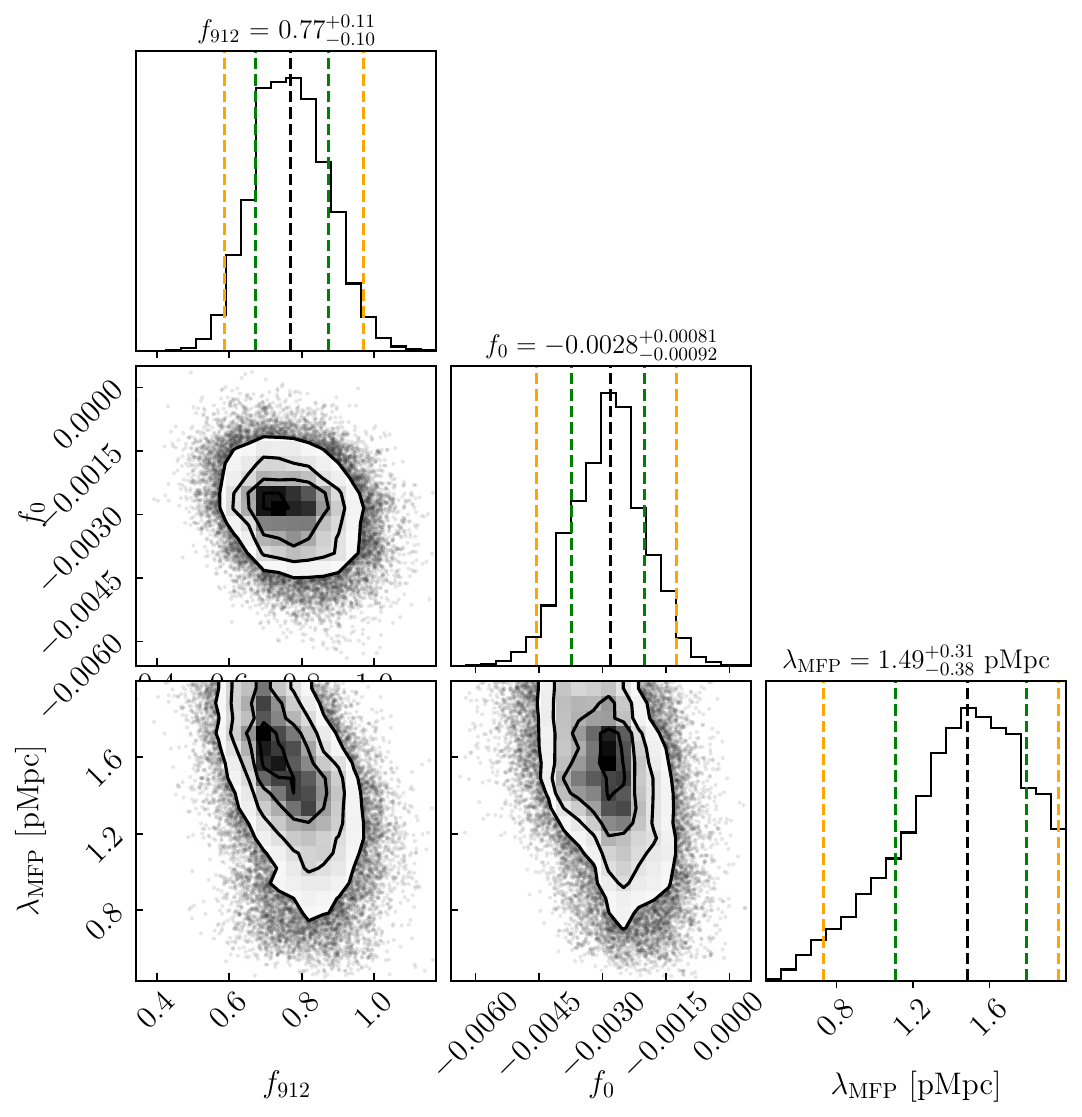}  
  \caption{Corner plot showing the posterior distributions of the normalization parameters ($f_{912}$, $f_{0}$) and the mean free path ($\lambda_{\rm{MFP}}$). The rightmost panels in each row indicate the marginalized posteriors. The black, green, and yellow dotted lines represent the median, 1$\sigma$, and $2\sigma$ levels of the distribution, respectively.}
  \label{fig:posterior}
\end{figure}

We assumed uniform priors in the ranges $0.3<$ \MFP\ $< 2$, $0< f_{912}< 1$ and $-0.001< f_{0}< 0.01$. The model stacks and the likelihood function were linearly interpolated over a grid of models wherever necessary.  We used the \texttt{emcee} package \citep{2013PASP..125..306F} to perform the inference by sampling the posterior distribution using MCMC via Bayes' theorem,
\begin{equation}
    p(\theta|\bf{d}) \propto \mathcal{L}(\bf{d}|\theta)p(\theta).
\end{equation}
The resultant best-fit curve, given by the mean of the posterior, is shown in Figure~\ref{fig:bestfit}, along with the 1$\sigma$ uncertainty. The marginalised posterior distributions of our model parameters is shown in Figure~\ref{fig:posterior}. 

We expect our assumptions for quasar lifetimes and host halo masses to introduce additional uncertainty on these parameters.  We approximate these uncertainties by adding them in quadrature to the uncertainty derived from the posterior distributions.  We saw in Figure~\ref{fig:mfppdf_tqhm} that varying quasar lifetimes and host halo masses leads to a $\sim$10\% and $\sim$25\% change, respectively, in the mean free path. With this additional error included, our inferred value of MFP is $\lambda_{\mathrm{MFP}}=1.49^{+0.47}_{-0.52}$~pMpc.

\begin{figure*}
  \includegraphics[width=0.8\textwidth]{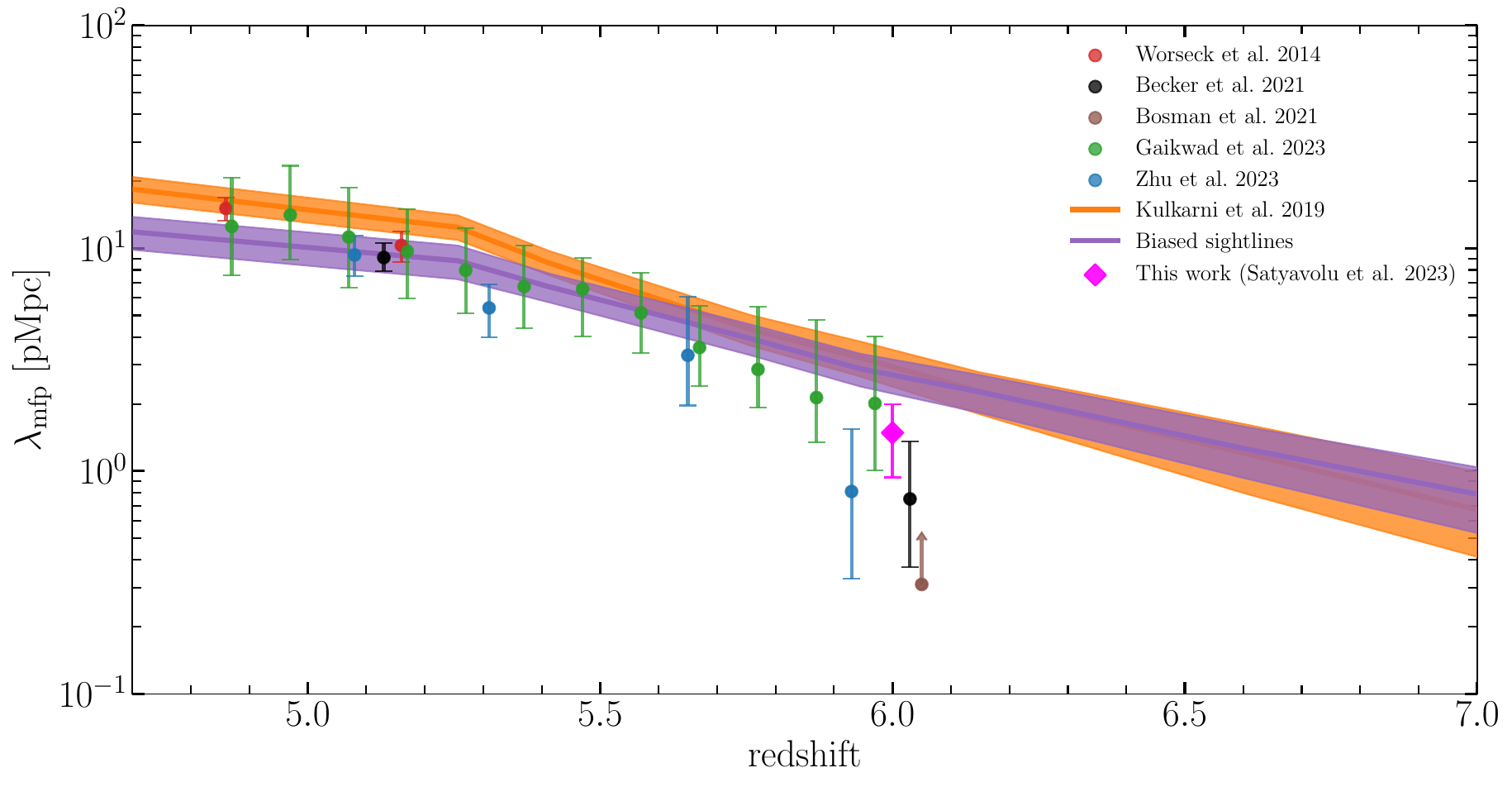}
  \caption{The magenta point shows our measurement of the MFP using the \beck\ data with our model.  The error bar on this point shows the 1$\sigma$ uncertainty.  We see that this measurement is consistent with that of \beck\, which is shown by the black point.  This suggests that the \beck\ measurement is robust against the sources of potential bias that we have considered in this paper.  The orange curve with shaded regions shows the mean free path in the reionization simulations of \citet{2019MNRAS.485L..24K}.  We see that our new measurement prefers a factor-of-two smaller MFP than this simulation.  We also show other measurements of the MFP, by \citet{2014MNRAS.445.1745W}, \citet{2021arXiv210812446B}, \citet{2023arXiv230402038G}, and \citet{2023ApJ...955..115Z}. Measurements by \citet{2021MNRAS.508.1853B,2023arXiv230402038G} and \citet{2021arXiv210812446B} have been displaced in redshift by $\delta z=\pm$  0.03 and 0.05 respectively for legibility.}
  \label{fig:ourmfp_v_sim}
\end{figure*}

Figure~\ref{fig:ourmfp_v_sim} shows the comparison of our new measurement with previous measurements and our simulation. While our new measurement is slightly higher than the MFP measured by \beck, it is consistent with their result. Our measurement is also consistent with the other estimates by \citet{2023arXiv230402038G} and \cite{2023arXiv230804614Z}.  Our new measurement remains lower than the MFP of our simulations by more than 2$\sigma$.  Nonetheless, the consistency between our inference and that of \beck\ suggests that the \beck\ method is robust with respective to the quasar proximity zone modelling.

\section{Conclusions}\label{sec:conclusion}

In this paper, we have taken a closer look at the mean free path of hydrogen-ionizing photons at $z=6$ in order to examine potential sources of bias in recent measurements.  Our findings are as follows:

\begin{itemize}
\item At least for $z\lesssim 6.5$, there is no significant difference in the value of the mean free path obtained by using any of the multiple definitions of this quantity that have been used in the recent literature for redshifts at which reionization is still incomplete.
\item The bias in the mean free path due to the overdensities around quasars is also minimal at $z\sim 6$.  At lower redshifts, with $z\lesssim 5$, the overdensities can bias the mean free path towards lower values by about 50\%.  
\item Due to the short mean free path at $z\sim 6$, the \lyc\ composite spectra used in the literature for a direct MFP measurement are affected by the quasar proximity zones.  The dependence of the inferred mean free path on the variations in quasar lifetime and host halo masses appears, however, only to be small and less than 25\%.
\item The patchiness of reionization also has minimal effect on the direct measurement of the MFP reported by \beck.
\item Using radiative transfer modelling of patchy reionization and quasar proximity zones, we reanalyze the data obtained by \beck\ to find a MFP of $\lambda_{\mathrm{MFP}}=1.49^{+0.47}_{-0.52}$~pMpc at $z=6$. This is nearly two times larger than the \beck\ measurement and smaller than the MFP in our reionization simulation by less than a factor of $\sim 2$.
\end{itemize}

We fit to the data while self-consistently modelling quasar proximity effect on the MFP in our simulations. We include the quasars through post-processing and hence do not consider the response of gas densities to the photoheating caused by the quasar ionization. However, the dynamical timescale for the gas to respond ($\sim 100$ Myr) is usually much larger  than the average episodic lifetime of the quasar, which is around $\sim 1$~Myr \citep{2020ApJ...898..149D,2021ApJ...921...88M,2023MNRAS.521.3108S}.  We also do not include feedback from AGN which could result in gas heating as well as AGN driven clumpy outflows, although the former effect is shown to be not major \citep{2018MNRAS.478.5607T}.  The response of gas densities to heating from reionization might nonetheless be not captured in our main cosmological RT simulation, which was also run in post-processing. This would require running either hybrid or fully coupled cosmological hydrodynamical simulations \citep[e.g.][]{2023MNRAS.519.6162P,2022MNRAS.512.4909G}.  We did not check the dependence of our result on the quasar spectral profile due to computational costs. The error on MFP due to this however is expected to be small (\beck).

\citet{2021ApJ...917L..37C} have modeled sinks in a subgrid fashion and found that the small-scale clumping reduces the MFP.  There has also been evidence for presence of excess LLSs along the line-of-sight to quasars \citep{2006ApJ...651...61H}.  We furthermore do not resolve the dense gas within mini-halos in our simulation. Recent works show that the \lya\ flux is suppressed by 10\% on average due to mini-halos, with the suppression being enhanced along lines of sight in the vicinity of large halos, above redshifts $z\gtrsim 5.5$ \citep{2023arXiv230904129P}.  Inclusion of dense absorbing gas such as LLSs or mini-halos in our simulations therefore will not only reduce the MFP along the random sightlines shown in orange in Figure~\ref{fig:mfpsim_v_obs}, but also further reduce the continuum flux along quasar sightlines. This would push our measured MFP in Figure~\ref{fig:ourmfp_v_sim} to higher than its current value.

In summary, the shortness of the mean free path relative to radiative transfer simulation could be due to unresolved excess Lyman-limit systems (LLSs) along quasar sightlines.  Consequently, the overall picture that emerges is consistent with one in which reionization ends much later than redshift 6, with the photon budget dominated by faint galaxies with high ionizing efficiency \citep{2021ApJ...918L..35D, 2021ApJ...917L..37C}.

\section*{Acknowledgements}
We thank George Becker and Yongda Zhu for providing data and for valuable discussions. We also thank James Bolton for sharing Sherwood data.
We thank Shikhar Asthana, Anson D'Aloisio, Christopher Cain, Prakash Gaikwad, Joe Hennawi, Vid Iršič and Nabendu Kumar Khan for useful discussions.  GK is partly supported by the Department of Atomic Energy (Government of India) research project with Project Identification Number RTI~4002, and by the Max Planck Society through a Max Planck Partner Group.  This work was supported by grants from the Swiss National Supercomputing Centre (CSCS) under project IDs s949 and s1114. This work used the Cambridge Service for Data Driven Discovery (CSD3) operated by the University of Cambridge (www.csd3.cam.ac.uk), provided by Dell EMC and Intel using Tier-2 funding from the Engineering and Physical Sciences Research Council (capital grant EP/P020259/1), and DiRAC funding from the Science and Technology Facilities Council (www.dirac.ac.uk).  This work further used the COSMA Data Centric system operated Durham University on behalf of the STFC DiRAC HPC Facility. This equipment was funded by a BIS National E-infrastructure capital grant ST/K00042X/1, DiRAC Operations grant ST/K003267/1 and Durham University. DiRAC is part of the UK's National E-Infrastructure. Support by ERC Advanced Grant 320596 ‘Emergence’  is gratefully acknowledged. For the purpose of open access, LK has applied a Creative Commons Attribution (CC BY) licence to any Author Accepted Manuscript version arising from this submission. 

\section*{Data Availability}

The data and code underlying this article will be shared on reasonable request to the corresponding author.

\bibliographystyle{mnras}
\bibliography{refs} 

\bsp	
\label{lastpage}
\appendix 
\section{Distribution of free paths}\label{sec:appendixa}
\begin{figure}
    \centering
    \includegraphics[width=\columnwidth]{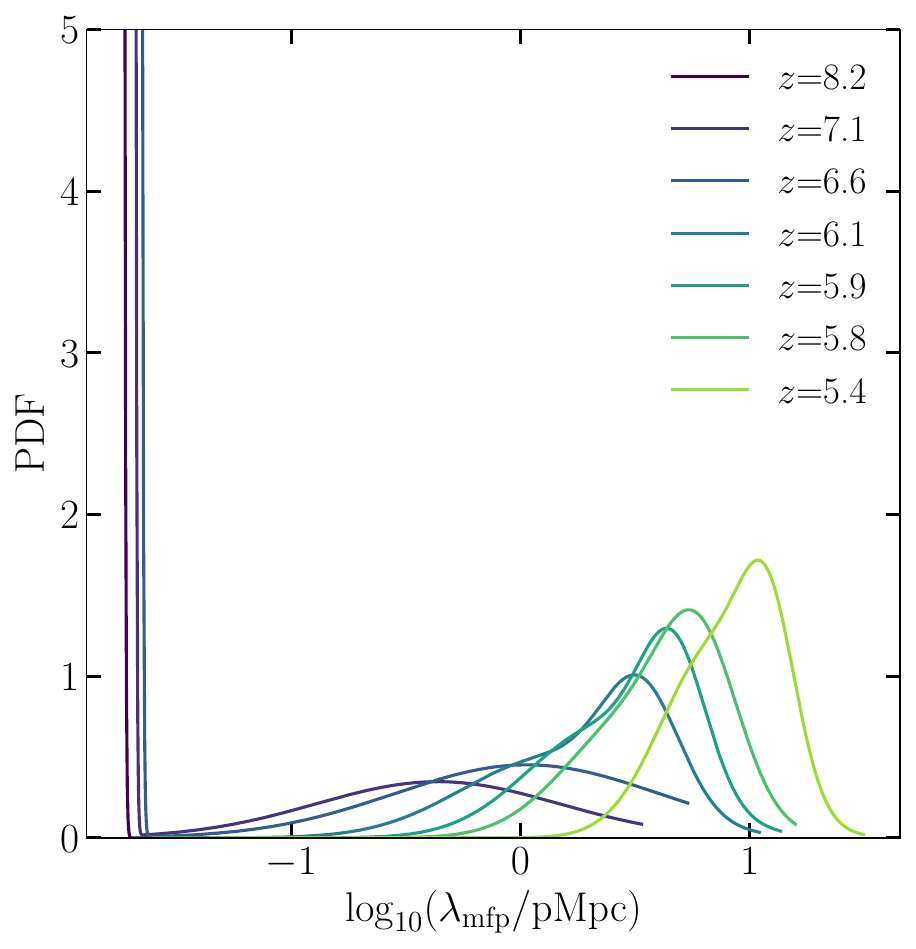}
    \caption{Distribution of free paths across redshifts. At redshifts $z<6$, the free path length distribution becomes bimodal with a large number of free paths $<0.01$~pMpc, smaller than the resolution of our base simulation at these redshifts.}
    \label{fig:bimodality}
\end{figure}

Figure~\ref{fig:bimodality} shows the distribution of free paths in our simulation between redshifts 8 and 5. The divergence seen at redshifts $z<6$ are numerical artefacts arising from the short free paths limited by the resolution of our simulation. 

\section{Effect of off-diagonal terms in the covariance matrix} 

\begin{figure}
    \centering
    \includegraphics[width=\columnwidth]{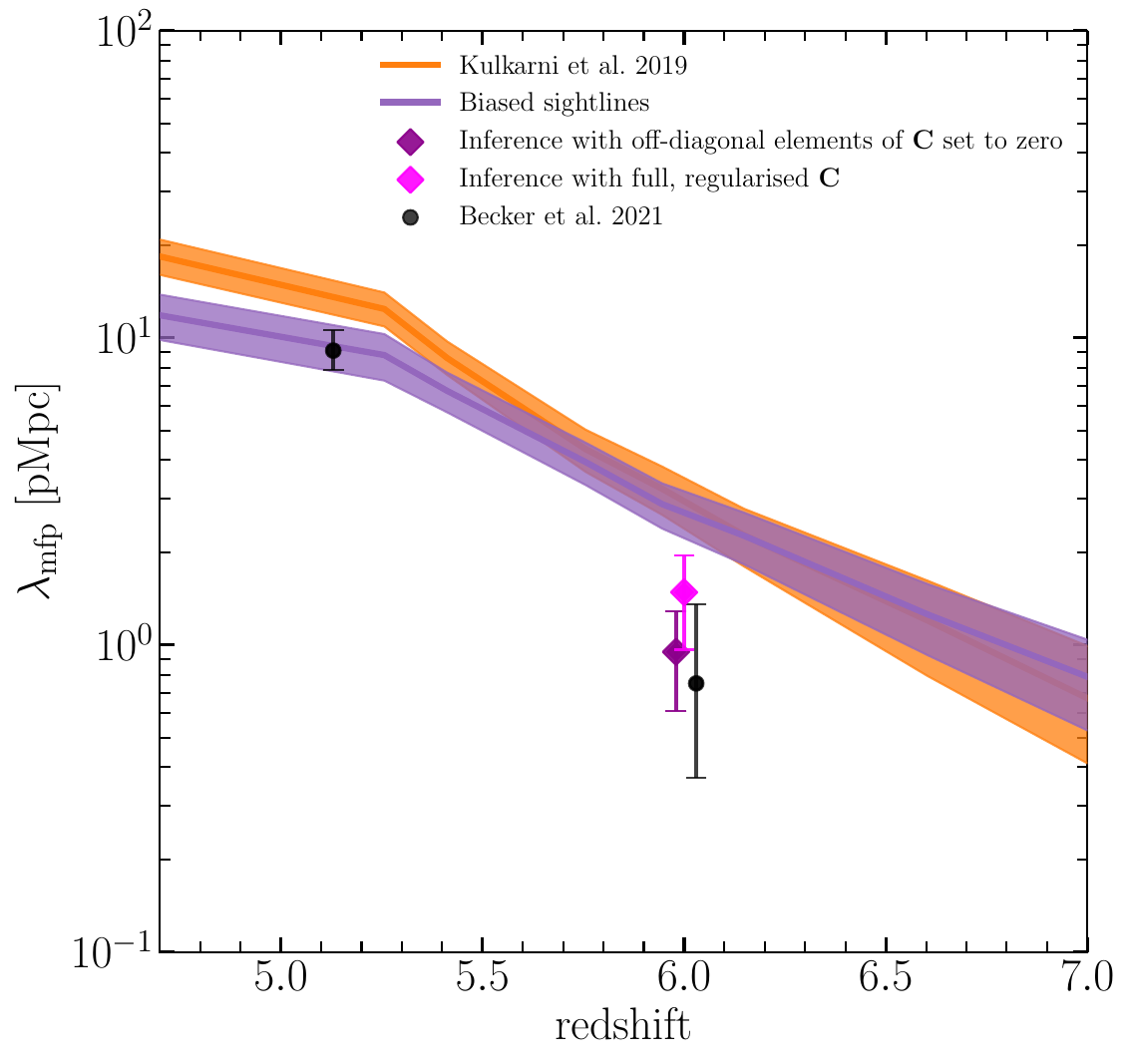}
    \caption{Effect of the off-diagonal terms in the covariance matrix $\bf{C}$ on our inferred value of the mean free path.  Magenta diamond shows the mean of the posterior distribution computed using the full, regularised covariance mastrix as described in the text.  Also shown is the one-sigma uncertainty.  The purple diamond shows the same for the case in which only the diagonal elements of the covariance matrix are used while computing the likelihood.  In this case the off-diagonal terms are set to zero. The black points show the \beck\ measurements.  We see that the off-diagonal elements lead to a larger mean value of the posterior distribution and a narrower posterior distribution, resulting in lower 1$\sigma$ uncertainty.  The three MFP measurements shown at $z\sim 6$ are at $z=6$ but we have shown them with small displacements in the redshift directions for better legibility.}
    \label{fig:mfp_covcompare}
\end{figure}

To understand the effect of the off-diagonal terms in the covariance matrix, we repeat our analysis of Section~\ref{sec:results} by fitting our model to the stacked data of \beck\ using only the diagonal elements of the data covariance matrix provided by \beck. The full, unregularised, data covariance matrix is nearly diagonal, but has some non-zero off-diagonal elements, particularly around $\sim912$\AA. Setting all off-diagonal terms to zero results in an inference shown by the purple diamond in Figure~\ref{fig:mfp_covcompare}.  The magenta diamond in this figure, on the other hand, shows the result of the analysis from Section~\ref{sec:results}.  We see that the off-diagonal elements lead to a larger mean value of the posterior distribution and a narrower posterior distribution, resulting in lower 1$\sigma$ uncertainty. 


\end{document}